\documentclass[CRPHYS,Unicode,manuscript]{cedram}

\usepackage{graphicx}
\usepackage{dcolumn}
\usepackage{bm}
\usepackage{amsfonts}
\usepackage{amssymb}
\usepackage{comment}

\newcommand{\bfq}{\mathbf{q}}
\newcommand{\bfR}{\mathbf{R}}
\newcommand{\bfr}{\mathbf{r}}
\newcommand{\bfd}{\mathbf{d}}
\newcommand{\bfzo}{\mathbf{0}}
\newcommand{\bfx}{\mathbf{x}}
\newcommand{\bfk}{\mathbf{k}}

\newcommand{\clO}{\mathcal{O}}
\newcommand{\clC}{\mathcal{C}}
\newcommand{\clS}{\mathcal{S}}
\newcommand{\clZ}{\mathcal{Z}}
\newcommand{\clP}{\mathcal{P}}

\newcommand{\LLL}{\mathrm{LLL}}
\newcommand{\rmH}{\mathrm{H}}
\newcommand{\intt}{\mathrm{int}}
\newcommand{\rel}{\mathrm{rel}}
\newcommand{\tr}{\mathrm{tr}}
\newcommand{\eff}{\mathrm{eff}}
\newcommand{\rmB}{\mathrm{B}}
\newcommand{\rmL}{\mathrm{L}}
\newcommand{\rmr}{\mathrm{r}}

\newcommand{\const}{\mathrm{const}}
\newcommand{\com}{\mathrm{com}}
\newcommand{\eqr}{\mathrm{eq}}

\graphicspath{{./figures/}}

\let\oldtilde\tilde
\renewcommand*{\tilde}[1]{\mathchoice{\widetilde{#1}}{\widetilde{#1}}{\oldtilde{#1}}{\oldtilde{#1}}}
\let\oldhat\hat
\renewcommand*{\hat}[1]{\mathchoice{\widehat{#1}}{\widehat{#1}}{\oldhat{#1}}{\oldhat{#1}}}

\RubricEF{Gold Medal Jean Dalibard}{Médaille d'or Jean Dalibard}

\title{Scale Invariance in the Lowest Landau Level}
\alttitle{Invariance d'échelle dans le niveau de Landau fondamental}

\author{\firstname{Johannes} \lastname{Hofmann}\IsCorresp}
\email[J. Hofmann]{johannes.hofmann@physics.gu.se}
\address{Department of Physics, Gothenburg University, 41296 Gothenburg, Sweden}
\thanks{This work is supported by Vetenskapsr\aa det (grant number 2020-04239).}

\author{\firstname{Wilhelm} \lastname{Zwerger}\IsCorresp}
\email[W. Zwerger]{zwerger@tum.de}
\address{Technische Universit\"at M\"unchen, Physik Department, James-Franck-Strasse, 85748 Garching, Germany}


\altkeywords{Condensats de Bose--Einstien en rotation rapide, niveau de Landau fondamental, développement du viriel, invariance d'échelle et anomalie d'échelle quantique, relations universelles/exactes de Tan}

\begin{abstract}
We show that the discrete set of pair amplitudes $A_m$ introduced by Haldane are an angular-momentum resolved generalization of the Tan two-body contact, which parametrizes universal short-range correlations in atomic quantum gases. The pair amplitudes provide a complete description of translation-invariant and rotation-invariant states in the lowest Landau level (LLL), both compressible and incompressible. To leading nontrivial order beyond the non-interacting high-temperature limit, they are determined analytically in terms of the Haldane pseudopotential parameters $V_m$, which provides a qualitative description of the crossover towards incompressible ground states for different filling factors. Moreover, we show that for contact interactions $\sim g_2 \delta^{(2)}(\bfx)$, which are scale invariant at the classical level, the non-commutativity of the guiding center coordinates gives rise to a quantum anomaly in the commutator $i [\hat{H}_{\LLL}, \hat{D}_R] = (2 + \ell \partial_\ell) \hat{H}_{\LLL}$ with the dilatation operator $\hat{D}_R$ in the LLL, which replaces the trace anomaly in the absence of a magnetic field. The interaction-induced breaking of scale invariance gives rise to a finite frequency shift of the breathing mode in a harmonic trap, which describes transitions between different Landau levels, the strength of which is estimated in terms of the relevant dimensionless coupling constant $\tilde{g}_2$.
\end{abstract}

\begin{altabstract}
Nous montrons que l'ensemble discret des amplitudes de paires $A_m$ introduit par Haldane est une généralisation résolue en moment cinétique du coefficient de contact à deux corps de Tan, qui paramétrise les corrélations universelles à courte distance dans les gaz quantiques atomiques. Les amplitudes de paires fournissent une description complète des états invariants par translation et par rotation dans le niveau de Landau fondamental (LLL), qu'ils soient compressibles ou incompressibles. Au premier ordre non nul au-delà de la limite de haute température sans interaction, elles sont déterminées analytiquement en fonction des paramètres $V_m$ du pseudopotentiel de Haldane, ce qui fournit une description qualitative du passage vers des états fondamentaux incompressibles pour différents taux de remplissage. De plus, nous montrons que pour les interactions de contact $g_2 \delta^{(2)}(\bfx)$, qui sont invariantes d'échelle au niveau classique, la non-commutation des coordonnées du centre de giration donne naissance à une anomalie quantique dans le commutateur $i [\hat{H}_{\LLL}, \hat{D}_R] = (2 + \ell \partial_\ell) \hat{H}_{\LLL}$ de l'hamiltonien avec le générateur des dilatations $\hat{D}_R$ dans le LLL, qui remplace l'anomalie de Weyl sur la trace en l'absence de champ magnétique. La brisure de l'invariance d'échelle induite par l'interaction conduit à un déplacement de fréquence du mode de respiration dans un piège harmonique, qui reflète des transitions entre différents niveaux de Landau et dont nous estimons la valeur en termes de la constante de couplage sans dimension pertinente $\tilde{g}_2$.
\end{altabstract}

\begin{DefTralics}
\newcommand{\bfk}{\mathbf{k}}
\newcommand{\LLL}{\mathrm{LLL}}
\end{DefTralics}

\begin{document}
\maketitle

\section{Introduction}

The observation by Tsui, St\"ormer and Gossard~\cite{tsui82} of a Hall conductance $\sigma_{\rmH}$ that is quantized at fractional values of the fundamental unit $\sigma_0=e^2/h$ in a high-mobility two-dimensional (2D) electron gas subject to a strong magnetic field $B$ has launched an immense amount of theoretical work dealing with interacting fermions in the lowest Landau level (LLL). In physical terms, the LLL is reached in the limit $\ell_B/a_0^{\eff}\to 0$ where the magnetic length $\ell_B=\sqrt{\hbar/eB}$ is much smaller than the effective Bohr radius $a_0^{\eff}=\hbar^2/m^*e^2$, where $e$ is the electron charge and $m^*$ its effective mass. Based on Laughlin's realization that for dominant short-range repulsion the ground state at filling factors $\nu=1/3, 1/5\ldots$ (or their particle-hole conjugates at filling $1-\nu$) forms an incompressible fluid with a finite excitation gap of order $\Delta\simeq e^2/\ell_B$~\cite{laughlin83}, an understanding of the quantization of $\sigma_{\rmH}$ that does not depend on any microscopic details is provided by an effective Chern-Simons field theory~\cite{froehlich91,wen92}. This description, which is based only on the underlying symmetries, has been extended by Son and coworkers~\cite{hoyos12,son13,son15}, which allows to properly account for effects like the Hall viscosity~\cite{hoyos12} or the important issue of particle-hole symmetry at half filling $\nu=1/2$~\cite{son15}.

While completely general, the effective-field-theory description of interacting particles in the LLL is restricted to incompressible ground states and it is a priori not clear whether further universal results can be derived beyond this special class of topologically ordered states. It is the aim of our present work to provide some steps in this direction, mostly restricting ourselves to the case of bosons in the context of ultracold quantum gases. In the special limit of vanishing interactions in the LLL, this problem can be mapped to a Gaussian field \mbox{$\psi(\bfx)=\sum_m a_m\phi_m(\bfx)$} whose expansion coefficients $a_m$ define a random polynomial. Its roots determine the location of vortices, which exhibit an antibunching property~\cite{castin06}. Here, we are instead concerned with the problem in the presence of interactions, which for Bose gases in 2D are usually described by an effective pseudopotential $V(\bfx)=g_2(\Lambda)\delta(\bfx)$~\cite{bloch08}. Since a delta function interaction does not give rise to a proper two-body scattering problem in two dimensions, the relevant strength $g_2(\Lambda)$ must be regularized by a logarithmic running with a cutoff~$\Lambda$. As discussed below, for states within the LLL, this renormalization turns out to be absent and \mbox{$g_2(\Lambda)\to (\hbar^2/m^*)\tilde{g}_2$} can be replaced by a dimensionless constant $\tilde{g}_2=\sqrt{8\pi}\, a/\ell_z$, which is determined by the 3D scattering length $a$ and the transverse confinement length $\ell_z$~\cite{bloch08}.

An effective magnetic field for neutral particles may be induced by rotating the gas with a finite angular frequency~$\Omega$~\cite{bloch08}. In the stationary rotating frame, the Hamiltonian $\hat{H}$ of the nonrotating system is changed to
\begin{equation}\label{eq:rotating-substitution}
\hat{H} \to \hat{H} - \Omega \hat{L}_z,
\end{equation}
where $\hat{L}_z$ is the angular momentum operator. This gives rise to a uniform effective magnetic field $(eB)_{\eff}=2m^*\Omega$ directed along the rotation axis $z$, reflecting the mathematical similarity between the Coriolis and Lorentz force (here, we keep $m^*$ as the atomic mass to avoid confusion with the angular momentum quantum number $m$ introduced below). Note that, in contrast to the electronic problem, the effective magnetic length $\ell_B = \sqrt{\hbar/(eB)_{\eff}} = \sqrt{\hbar/2m^*\Omega}$ now scales inversely with the mass $m^*$, while the effective cyclotron frequency $\omega_c = 2 \Omega$ is independent of it. In the presence of an additional harmonic confinement with frequency $\omega$, the single-particle levels in the $x,y$-plane are then of the form~\cite{bloch08}
\begin{equation}\label{eq:single-particle}
E_j^{(n)}=
\hbar(\omega-\Omega)\cdot j + \hbar(\omega+\Omega)\cdot n\,,
\end{equation}
where we drop a constant zero-point energy $\hbar \omega$. In the limit $\Omega\to\omega^{-}$ and for given $n=0,1,\,\ldots$, these energies group into a series of degenerate levels labelled by their angular momentum $m=j-n=-n,-n+1\ldots$, which constitute an analog of the $n^{\rm th}$ Landau level. In particular, the LLL corresponds to $n=0$ with a degenerate set of single-particle levels labelled by the angular momentum quantum number $m=j \geq 0$. In the following, we use as a characteristic length scale the harmonic oscillator length
\begin{equation}
\ell = \sqrt{\frac{\hbar}{m^* \omega}},
\end{equation}
which differs by a factor of $\sqrt{2}$ from the effective magnetic length $\ell_B$ in the LLL limit. Quite generally, the restriction to the LLL requires that $|V_m|/\hbar\omega\to 0$ in order to suppress transitions between different Landau levels, where $|V_m|$ is the characteristic magnitude of the Haldane pseudo\-potentials formally introduced in Eq.~\eqref{eq:Vm} below. For bosons, the dominant interaction is $V_0=\hbar\omega\cdot \tilde{g}_2/2\pi$~\cite{cooper08}, which arises from the standard zero-range pseudopotential with scattering length $a$ discussed above. Exceptions arise, however, if the short-range scattering length is tuned to zero or in the presence of dipolar interactions, where---in a minimal model---both~$V_0$ and~$V_2$ need to included~\cite{cooper05}. Note that since the dimensionless parameter $\tilde{g}_2$ parametrizes an $s$-wave interaction, it takes the same form for any rotation frequency and is thus independent of the magnetic length~$\ell_B$, cf.~App.~\ref{app:twobody}. The condition $V_0/\hbar \omega \sim \tilde{g}_2 \ll 1$ for staying in the LLL corresponds to the perturbative limit with respect to the harmonic oscillator spacing, for which the running of the interaction strength due to the renormalization of the coupling is negligible. Since $\tilde{g}_2\lesssim 0.1$ generically, the LLL condition is well obeyed in practice\footnote{Note that the relevant dimensionless interaction strength \emph{within} the LLL is set by $V_0/\hbar(\omega-\Omega)$ and thus becomes strong for $\Omega \to \omega^-$.}.

For bosons, the filling factor $\nu=n_2\,(\pi\ell^2)$ in the LLL ($n_2$ is the 2D particle density) may take arbitrary large values. In particular, for a gas with $N$ particles in a rotating harmonic trap~\cite{bloch08},
\begin{equation}\label{eq:filling}
\nu\simeq \left(N(1-\Omega/\omega)/\tilde{g}_2\right)^{1/2}
\end{equation}
scales like $\sqrt{N}$ at a fixed value of $1-\Omega/\omega\ll 1$. In the regime of large filling factors $\nu\gg 1$, which is the one that has been explored mostly so far, bosons can be described in terms of a coherent state picture. In particular, the ground state is a regular array of vortices, which is predicted to melt into a strongly correlated vortex liquid at around $\nu\simeq 10$~\cite{cooper01} (for a review, see Ref.~\cite{cooper08}). From Eq.~\eqref{eq:filling}, the condition for reaching filling factors of order unity requires $1-\Omega/\omega\lesssim \tilde{g}_2/N$, which places the gas very close to an unstable configuration where the harmonic trap can no longer overcome the centrifugal force. As a result, a number of alternative routes have been proposed to realize Bose gases in the LLL in a regime where mean-field theory no longer applies, such as optical flux lattices~\cite{cooper11,cooper11b} or periodically driven systems, where a nonvanishing gauge field arises in the effective Floquet Hamiltonian~\cite{goldman14, eckardt17}. This idea has been implemented successfully in a recent experiment by L\'eonard et al~\cite{leonard22}, where a strongly correlated state with $\nu=1/2$ has been observed with $N=2$ bosons in an optical lattice. Earlier, the physics of strongly interacting bosons in the LLL has also been realized with photons in a twisted optical cavity, where the repulsion results from an effective interaction mediated by Rydberg atoms~\cite{clark20}. Moreover, a quite different approach to study interacting bosons in the LLL has recently been explored with $^{23}$Na atoms that are geometrically squeezed into the LLL using a rotating saddle trap~\cite{fletcher21,mukherjee22}.

This paper is structured as follows: In Sec.~\ref{sec:pairamplitudes}, we show that the pair amplitudes $A_m$ originally introduced by Haldane~\cite{haldane90} via the number of particle pairs with relative angular momentum $m$ may be seen as a generalization of the Tan contact parameter for non-rotating quantum gases with zero-range interactions. The pair amplitudes fully describe fluid states within the LLL at arbitrary temperature and filling. An explicit result for the $A_m$ as a function of the Haldane pseudo\-potential parameters $V_m$ is derived within a virial expansion. It turns out that, even at leading order, this allows to describe in qualitative terms the smooth crossover towards incompressible ground states, with a non-monotonic behavior of the compressibility as a function of temperature. Moreover, data for the pair distribution function obtained in the recent experiment by L\'eonard et al.~\cite{leonard22} allows to extract the two lowest Haldane pair amplitudes $A_0$ and $A_2$ in the strongly correlated state at $\nu=1/2$. In Sec.~\ref{sec:scaleinvariance}, we consider the fate of scale invariance and its violation by quantum fluctuations of 2D Bose gases in the presence of rotation. It is shown that the interaction-induced quantum scale anomaly of the nonrotating system is suppressed in the LLL limit due to the absence of a running coupling constant. Nevertheless, the frequency of the breathing mode, which here describes transitions between different Landau levels, is shifted away from the symmetry-dictated value. In addition, we show that despite the absence of a running coupling constant, a quantum anomaly is still present in the LLL due to the non-commutative nature of the guiding center coordinates. The paper ends with a conclusion in Sec.~\ref{sec:conclusion} and contains two appendices that summarize the two-body problem in a rotating trap and provide details on the virial expansion in the LLL, respectively.

\section{Haldane pair amplitudes and Tan-like relations in the LLL}\label{sec:pairamplitudes}

In order to describe interactions in the LLL, we start from the general interaction Hamiltonian
\begin{equation}\label{eq:ham1}
\hat{H}_{\intt} = \frac{1}{2} \int\!dz_1\!\int\! dz_2 \, V(z_1-z_2) \, \hat{\psi}^\dagger(z_2) \hat{\psi}^\dagger(z_1) \hat{\psi}(z_1) \hat{\psi}(z_2)
\end{equation}
with a translation invariant two-body interaction $V(z)$ in the absence of rotation. Here, $z=x+iy$ is a complex coordinate and $\hat{\psi}^\dagger(z)$ is the creation operator for a particle at position $z$. To incorporate the finite rotation $\Omega \to \omega^-$, where the kinetic energy is completely quenched, as well as the restriction to the LLL, it is convenient to consider a disk geometry in the circular gauge. Expanding the field operator $\hat{\psi}_{\LLL}(z) = \sum_m \phi_m(z) \hat{a}_m$ restricted to the LLL in terms of the associated single-particle eigenstates $\phi_m(z) = (z/\ell)^{m} \, e^{-\bar{z} z/2\ell^2}/\sqrt{\pi \ell^2 m!}$, the projected interaction Hamiltonian is given by
\begin{equation}\label{eq:HLLL}
\hat{H}_{\intt}^{\LLL} = {\sum_{m}}^{'} V_m\, \sum_{M} \, \hat{\xi}_{mM}^\dagger \hat{\xi}_{mM}^{} \; = \;{\sum_{m}}^{'} V_m\, \hat{P}_{m}\,.
\end{equation}
Here, the projection operator \mbox{$\hat{P}_{m}=\sum_M \, \hat{\xi}_{mM}^\dagger \hat{\xi}_{mM}$} is defined in terms of the operator
\begin{equation}
\hat{\xi}_{mM} = \frac{1}{\sqrt{2}} \sum_{m_1,m_2} \left\langle mM \,\middle|\, m_1, m_2 \right\rangle \hat{a}_{m_1} \hat{a}_{m_2},
\end{equation}
which annihilates a two-particle state $|mM\rangle$ described by the center-of-mass quantum number $M$ and a relative angular momentum $m$. The prime in Eq.~\eqref{eq:HLLL} restricts the $m$-summation to odd or even non-negative integers in the case of fermions or bosons, respectively. In the presence of both translation and rotation invariance, there is no dependence on the center-of-mass quantum number $M$, and the two-body interaction $V(z)$ within the LLL reduces to a discrete set of Haldane pseudopotential parameters~\cite{haldane83}
\begin{equation}\label{eq:Vm}
V_m = \left\langle mM \left| \hat{V}\right| mM \right\rangle \;\; \xrightarrow[m\,\gg\,1] \;\; V\left(z\!=\!\ell\sqrt{2m}\right)\,.
\end{equation}
Specifically, for a zero-range pseudopotential, only $V_0$ is non-vanishing as discussed above, while for Coulomb interactions the $V_m$ decay rather slowly as $1/\sqrt{m}$ for $m\gg 1$. The projection operators $\hat{P}_m=N_m\,\hat{A}_m$ are determined by a degeneracy factor $N_m$, which counts the number of allowed values of the quantum number $M$, and the $M$-independent pair amplitude operators
\begin{equation}\label{eq:pair-amplitudes}
\hat{A}_m = \hat{\xi}_{mM}^\dagger \hat{\xi}_{mM}^{},
\end{equation}
which were first introduced by Haldane~\cite{haldane90}. Their expectation values $A_m=\langle\hat{A}_m\rangle$ are positive by definition and they depend on the complete set $\{ V_m\}$ of the Haldane pseudoptential parameters as well as the specific state in question. For $m\gg 1$, they approach a trivial constant $A_m\to\nu^2$ since the expectation value factorizes. In the case of bosons, the $A_m$ may be arbitrarily large, while for fermions, they are bounded from above by $A_m\leq 1$.

The pair amplitudes defined in~\eqref{eq:pair-amplitudes} provide a generalization of a concept introduced by Tan~\cite{tan08a,tan08b,tan08c} and by Zhang and Leggett~\cite{zhang09} in the context of ultracold gases, for which the interactions are replaced by a pseudopotential that is proportional to the exact scattering length $a$ of the underlying microscopic two-body potential~$V(\bfx)$. Tan derived a set of exact results valid at all interaction strengths that describe thermodynamic properties as well as the non-analytic short-distance and short-time structure of the system, which in turn characterize the high-momentum and high-frequency tails of the system's response~\cite{braaten12}. The central quantity in all of these relations is the intensive Tan two-body contact density $\clC_2$. In 2D, at finite temperature, it is defined by the derivative
\begin{equation}\label{eq:contact-definition}
\frac{\partial f}{\partial\left(\ln{\, a_2}\right)}=\frac{\hbar^2}{4\pi m^*} \, \clC_2
\end{equation}
of the free energy density $f$ with respect to the logarithm of the 2D scattering length $a_2$, which in turn is related to its 3D counterpart $a$ via $-\ln{\, a_2}= \sqrt{\pi/2}\ell_z/a +\const$~\cite{bloch08}. The contact density determines the singular behavior $g^{(2)}(r) \to (\clC_2/(2\pi n_2)^2)\, \ln^2{r}$ of the pair distribution function at short distances and thus describes the anomalous enhancement to detect two particles in close proximity in the presence of zero-range interactions~\cite{braaten12}. Moreover, it sets the strength of the asymptotic power law $n(\bfk)\to \clC_2/k^4$ of the momentum distribution $n(\bfk)$, characteristic for zero-range interactions in any dimension\footnote{Note that no such power laws appear in the LLL due to the analytic nature of the associated many-body wave functions.}. Specifically, one obtains $\clC_2=(n_2\tilde{g}_2)^2$ in the ground state of a weakly interacting Bose gas in 2D. In a similar manner, a set of extensive but now dimensionless ``contact coefficients'' may be defined for the many-body problem in the LLL by
\begin{equation}\label{eq:contacts}
C_2^{(m)}\left(\{V_m\}\right) = \frac{\partial F}{\partial V_m} = \left\langle \hat{P}_m\right\rangle = \sum_M \,\left\langle \hat{\xi}_{mM}^\dagger \hat{\xi}_{mM}^{} \right\rangle = N_m\cdot A_m\,,
\end{equation}
where $F=fA$ is the free energy with $A$ the system area and $\langle\ldots\rangle$ denotes a thermal average with a statistical operator $\hat{\rho}\sim\exp{(-\beta\hat{H}_{\intt}^{\LLL})}$. In physical terms, the dimensionless coefficients $C_2^{(m)}$ are just the expected number of particle pairs with relative angular momentum $m$. They determine the internal interaction energy $U=\langle\hat{H}_{\intt}^{\LLL}\rangle=\sum_m V_{m} C_2^{(m)}$ and may be viewed as an angular-momentum resolved generalization of the Tan two-body contact. An immediate consequence of the definition~\eqref{eq:contacts} is the sum rule
\begin{equation}
{\sum_{m}}^{'} C_2^{(m)} = {\sum_{m}}^{'} N_m \left\langle \hat{A}_m \right\rangle = \frac{N(N-1)}{2},
\end{equation}
which counts the total number of pairs for fixed finite particle number $N$. In the limit $N\gg 1$, using the asymptotic result $N_m\to 2N/\nu$ for the degeneracy factor of a disc, the intensive contact densities $\clC_2^{(m)}=C_2^{(m)}/N$ are connected with the pair amplitudes by the simple relation
\begin{equation}\label{eq:contacts2}
\clC_2^{(m)}=\frac{2}{\nu} A_m \,\to\, 2\nu \quad \text{for} \quad m\to\infty,
\end{equation}
a result which holds for both bosons and fermions.

The discrete set of pair amplitudes $A_m$ or the related contacts $\clC_2^{(m)}$ provide a complete description of translation and rotation invariant many-body states in the LLL. In particular, they fully determine the associated pair distribution function
\begin{equation}\label{eq:pc}
n(z_1) n(z_2) g^{(2)}(z_1, z_2) = 2 \sum_{mM} \sum_{m'M'} \left\langle mM\,\middle|\, z_1 z_2 \right\rangle \left\langle z_1 z_2\,\middle|\, m'M' \right\rangle \left\langle \hat{\xi}_{mM}^\dagger \hat{\xi}_{m'M'}^{} \right\rangle.
\end{equation}
Indeed, rotational invariance implies $m=m'$ and $M=M'$. Furthermore, the condition that any dependence on the center of mass coordinate $Z$ disappears requires that $\langle \hat{\xi}_{mM}^\dagger \hat{\xi}_{mM}^{} \rangle$ does not depend on $M$. The summation over $M$, which runs from $M=0,\,\ldots,\,\infty$ in the thermodynamic limit $N\to\infty$, then just yields a factor of $N_m$ and the pair distribution function reads
\begin{equation}\label{eq:guniform}
g^{(2)}(z) = \frac{2}{n^2} \sum_{mM} \left\langle mM\,\middle|\, z_1 z_2 \right\rangle \left\langle z_1 z_2\,\middle|\, mM \right\rangle \, A_m = \frac{1}{\nu^2} \sum_{m} {}^{'} \frac{2}{m!} \left(\frac{|z|^2}{2 \ell^2}\right)^{m} e^{- |z|^2/2\ell^2} \, A_m\,,
\end{equation}
a representation that was first derived by Girvin~\cite{girvin84} for fluid ground states of fermions, where only odd values of $m=1,3,\,\ldots$ appear. For bosons, where the summation runs over even $m=0,2\ldots$, the pair distribution function approaches a finite value $g^{(2)}_{\,\rmB}(0)=2A_0/\nu^2$ at vanishing distance, which is trivially connected to the intensive contact density $\clC_2=\clC_2^{(m=0)}$ for vanishing relative angular momentum by $\clC_2=\nu\cdot g^{(2)}_{\,\rmB}(0)$. The contact coefficients for general values of $m$ can be expressed in terms of moments of the static structure factor $S(\bfq)$ by using the relation~\cite{haldane11}
\begin{equation}\label{eq:contact-density}
\clC_2^{(m)} = \frac{\left\langle \hat{P}_m\right\rangle}{N} = 2 \nu + 2\pi\ell^2 \int\frac{d^2q}{(2\pi)^2} L_m\left(q^2 \ell^2\right) \, \left[S(\bfq)-1\right],
\end{equation}
where $L_m(x)$ are Laguerre polynomials. Compared to the definition by Haldane in Ref.~\cite{haldane11}, we include in Eq.~\eqref{eq:contact-density} as the first term the Hartree contribution to the energy, which sets the large-$m$ behavior. Equation~\eqref{eq:contact-density} connects the $\clC_2^{(m)}$ to the static structure factor
\begin{equation}\label{eq:S(q)}
S(\bfq)= 1 + n_2 \int d^2x \, e^{-i\bfq\cdot\bfx}\,\left[g^{(2)}(\bfx)-1\right]=1-e^{-q^2\ell^2/2} + \bar{s}(\bfq)
\end{equation}
of the quantum fluid in the LLL. Here, $\bar{s}(\bfq)$ is the projected structure factor in the LLL~\cite{girvin86} and the subtraction in $S(\bfq)-1=\bar{s}(\bfq) - \exp{(-q^2\ell^2/2)}$ removes the diagonal self-interaction elements in the density summation. Remarkably, Eq.~\eqref{eq:contact-density} can be inverted, which allows to express the projected static structure factor
\begin{equation}\label{eq:barS}
\nu\bar{s}(\bfq)=\nu(1\pm \nu) e^{-q^2\ell^2/4} + 4 e^{-q^2\ell^2/2} \sum_m\,\!^{'} \tilde{A}_m\,L_m\left(\frac{q^2\ell^2}{2}\right)
\end{equation}
in terms of the connected pair amplitudes
\begin{equation}
\tilde{A}_m=A_m-\nu^2,
\end{equation}
where the upper or lower sign holds for bosons or fermions, respectively.

The introduction of connected pair amplitudes $\tilde{A}_m$, which decay to zero for $m\gg 1$, is useful for a number of reasons. First of all, decomposing $A_m=\nu^2+\tilde{A}_m$ guarantees that $g^{(2)}(\infty)=1$ for translation invariant states from the $m$-independent contribution $A_m^{(0)}=\nu^2$. More precisely, replacing $A_m$ by $\tilde{A}_m$ on the right-hand side of Eq.~\eqref{eq:guniform} gives the nontrivial interaction-dependent part $g^{(2)}(z)-(1\pm \exp{(-|z|^2/\ell^2)})$ of the pair distribution function in the LLL for bosons or fermions, respectively. A second, more important, point is that in the case of fermions at zero temperature, the connected pair amplitudes $\tilde{A}_m$ are particle-hole symmetric since the ground state and the resulting response and correlation functions like the combination $\nu\bar{s}(\bfq)$ are invariant under $\nu\to 1-\nu$~\cite{nguyen17} (for a detailed discussion of particle-hole symmetry in the LLL and beyond, see Ref.~\cite{zirnbauer21}).

A quite general result constraining the dependence of the contact coefficients on the set $\{ V_m\}$ of Haldane pseudopotentials follows from the fact that the free energy $F(\{ V_m\})$ must be a concave function of the coupling constants $V_m$. Indeed, the statistical operator $\hat{\rho}_{\eqr}\sim\exp{(-\beta\hat{H}_{\intt}^{\LLL})}$ in thermal equilibrium is a generalized Gibbs ensemble, which maximizes the entropy for given expectation values of the operators $\hat{P}_m$ with $\beta V_m$ as the associated Lagrange parameters. As a result, the symmetric matrix of derivatives
\begin{equation}\label{eq:concave}
\frac{\partial^2 F}{\partial V_m\,\partial V_n}=\frac{\partial C_2^{(m)}}{\partial V_n}
\end{equation}
is negative definite. In particular, for zero-range interactions where only the lowest Haldane pseudopotential parameter $V_0$ is present, the intensive contact $\clC_2=C_2^{(0)}/N=\nu \cdot g^{(2)}_{\,\rmB}(0)$ is a decreasing function of the interaction strength, e.g. $\partial\, \clC_2/\partial\, V_0<0$. In physical terms, this reflects the simple fact that the pair distribution function at vanishing separation decreases monotonically with increasing strength of the repulsion. In this context, it is also instructive to consider the special class of strongly correlated ground states whose pair amplitudes vanish identically below a given integer $q$, i.e. $A_m^{\rmL}=0$ for $m<q$, where $q=2,4\ldots$ in the case of bosons and $q=3,5\ldots$ for fermions~\cite{haldane11}. For these states, which include Laughlin's wave functions for the filling factors $\nu=1/q$ as a special case, the energy is independent of the corresponding Haldane pseudopotential parameters $V_{m<q}$. Moreover, as a consequence of the complete suppression of pairs with relative angular momentum $m<q$, the associated pair distribution function vanishes rather quickly like $g^{(2)}(z)\sim |z|^{2q}$ for small separations. While states with this property are usually considered as a paradigm for incompressible fluids in the LLL, it should be noted that the condition $\clC_2^{(m)}\equiv 0$ for $m<q$ is neither necessary nor sufficient for incompressibility. Indeed, incompressibility requires that the quadratic term $s_2$ in the expansion $\bar{s}(\bfq)=s_2 (q\ell)^2 + s_4 (q\ell)^4 +\ldots$ of the projected structure factor at zero temperature vanishes. As noted by Girvin, MacDonald and Platzman~\cite{girvin86}, this gives rise to a constraint on the pair amplitudes. In terms of their connected part, this can be written in the particle-hole symmetric form
\begin{equation}\label{eq:sumrule2}
\sum_m\,\!^{'}(m+1)\,\tilde{A}_m = - \frac{1}{8} \nu(1\pm \nu) \,,
\end{equation}
which is obtained directly by expanding Eq.~\eqref{eq:barS}. Obviously, fixing $\tilde{A}_m$ for just the lowest values of $m$ does not guarantee that this relation is obeyed.

\begin{figure}[t!]
\includegraphics{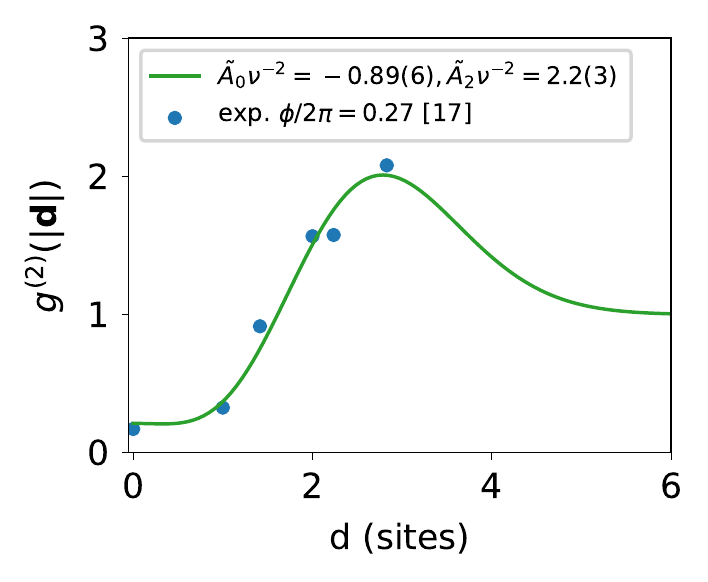}
\caption{The data for the angular averaged pair distribution function on a lattice from Ref.~\cite{leonard22} (blue dots) are compared with the exact result~\eqref{eq:guniform} for a continuum fluid of bosons in the LLL (green line). This allows to extract the two lowest Haldane amplitudes $A_0$ and $A_2$ for the strongly correlated state at $\phi/2\pi=0.27$, corresponding to a filling $\nu=1/2$. }
\label{fig:1}
\end{figure}

An example where the considerations above are of direct relevance is provided by the recent experiment by L\'eonard et al.~\cite{leonard22} with bosons in a small optical lattice in the presence of an effective magnetic field that corresponds to a filling factor $\nu=1/2$. In particular, the observed state with flux per unit cell $\phi/(2\pi)=0.27$ appears to be a few-body version of an incompressible fluid of bosons in the LLL. Experimentally, this interpretation is supported by using the Streda formula, which connects the Hall conductance $\sigma_H$ to the observable change in density with one in flux. As shown in Ref.~\cite{leonard22}, this gives $\sigma_{\rmH}/\sigma_0=0.6(2)$, consistent with the theoretical result $\sigma_{\rmH}/\sigma_0=\nu$ for a fractional Quantum Hall state. Now, the Streda formula relies on the assumption of incompressibility which is properly defined only at zero temperature and large particle numbers. A possible way to check whether the rather small experimental system on a $4\times 4$ lattice can indeed be considered as effectively incompressible is provided by an analysis of the pair distribution function measured in Ref.~\cite{leonard22}. Due to the small system size and the presence of an optical lattice, a comparison of the experimental results with Eq.~\eqref{eq:guniform} for the continuum problem in the thermodynamic limit is clearly only of a qualitative nature. Yet, it turns out that the azimuthal average (which restores rotation symmetry) of the measured data $g^{(2)}(\bfd)$ for the discrete lattice vectors $\bfd$ over a distance $r=|\bfd|$ between $r=0$ and about three lattice sites is sufficient to extract the two lowest Haldane amplitudes $A_0$ and $A_2$. This is shown in Fig.~\ref{fig:1}, where the data for the strongly correlated state at flux per unit cell $\phi/(2\pi)=0.27$ is compared with a fit to the exact result~\eqref{eq:guniform} for the continuum problem. The available data is described quite well by adjusting the values of the two lowest Haldane amplitudes $A_0$ and $A_2$ in Eq.~\eqref{eq:guniform}\footnote{It is important to note that the extracted values are essentially independent of the choice for the amplitudes with $m\geq 4$, because the single particle wave functions $\sim |z|^{2m}\exp{(-|z|^2/(2\ell^2))}$ in Eq.~\eqref{eq:guniform} are strongly localized near $|z|=r=\sqrt{2m}\,\ell$, which is beyond the range of the available data for $m\geq 4$. For the green line in Fig.~\ref{fig:1}, the associated Haldane amplitudes have been choosen to be $A_m=A_m^{(0)}=\nu^2$, which only affects the behavior beyond $|\bfd|\simeq 3$.}. In particular, the result $\tilde{A}_0/\nu^2=-0.89$ shows that $A_0=\nu^2(1+\tilde{A}_0/\nu^2)=0.028$ is close to zero, which is the value attained in a perfect Laughlin state at filling $\nu=1/2$ with many-body wave function $\psi^{\rmL}(z_1,\ldots z_N)\sim \prod_{i<j} (z_i-z_j)^2 \exp{(-\sum_i |z_i|^2/2\ell^2)}$. To demonstrate incompressibility as the crucial requirement for the quantization of $\sigma_{\rmH}$ of course requires large particle numbers $N\gg 1$ and a measurement of $g^{(2)}(r)$ for larger distances. This would give access to the higher-order Haldane pair amplitudes and thus a test of the criterion in Eq.~\eqref{eq:sumrule2}. Nevertheless, the data so far are consistent with a strongly correlated state of bosons in the LLL at small filling, which exhibits almost perfect anti-bunching at short distances.

The results above, some of which (like the representation~\eqref{eq:guniform} for the pair distribution function~\cite{girvin84} and the connection~\eqref{eq:contact-density} between the expectation values of the projectors $\hat{P}_m$ and the static structure factor~\cite{haldane11}) have been derived earlier, are completely general and they apply to fluid states in the LLL at arbitrary temperature. However, quantitative results for the pair amplitudes $A_m(\nu, \beta, \{V_m\})$ or contact coefficients for a given set $\{V_m\}$ of Haldane pseudopotentials can only be achieved numerically, e.g., by exact diagonalization for small particle numbers $N\lesssim 10$.\linebreak In the following, we show that a fully analytic result for the $A_m$ is available within a virial expansion. As derived in App.~\ref{app:virial}, the virial expansion in the LLL gives rise to a representation of the filling fraction in powers of the fugacity $z$ (not to be confused with a complex particle coordinate) of the form
\begin{align}
\nu &= \sum_{l=1}^\infty l b_l z^l = b_1 z + 2 b_2 z^2 + \ldots\label{eq:virialfilling}
\\
\intertext{with $b_1=1$ and}
b_2 &= \pm \frac{1}{2} - 2 \sum_m\,\!^{'} \left(1 - e^{-\beta V_m}\right),
\end{align}
where the positive sign holds for bosons and the negative sign for fermions. To leading order in the fugacity, the pair amplitude
\begin{align}
A_m &= \nu^2 e^{-\beta V_m} + {\clO}\left(z^3\right),\label{eq:virialcontact}
\\
\intertext{is identical for both cases and yields}
\tilde{A}_m &= \nu^2 \left(e^{-\beta V_m}-1\right) + {\clO}\left(z^3\right),\label{eq:virialAtilde}
\end{align}
for the connected pair amplitude. Obviously, the connected pair amplitudes vanish in the high-temperature limit $\beta V_m\to 0$, where $g^{(2)}(z)\to 1\pm \exp{(-|z|^2/\ell^2)}$ approaches the pair distribution function of a non-interacting gas of bosons or fermions in the LLL\footnote{Note that the result $g^{(2)}(z)=1+ \exp{(-|z|^2/\ell^2)}$ effectively describes bosons in the LLL at infinite temperature. This is quite different from the non-interacting gas at zero temperature considered by Castin et al.~\cite{castin06}, where $g^{(2)}(z)\sim |z|^2$ vanishes quadratically at short distances.}.
 A quite different high-temperature limit in the LLL has been discussed previously by Jeevanesan and Moroz~\cite{jeevanesan20} for the special case of bosons with zero range interactions. Using a Monte-Carlo approach for evaluating the classical grand canonical partition function, they have been able to determine the thermodynamics in the limit $V_0, |\mu| \ll T$, where the fugacity $z=\exp{(\beta\mu)}$ is of order one, over the full range of both negative and positive values of the scaling variable $x=\mu/\sqrt{V_0T}$, thus covering the crossover from a dilute vortex fluid to a vortex crystal.

\begin{figure}[!t]
\centering
\includegraphics[scale=0.95]{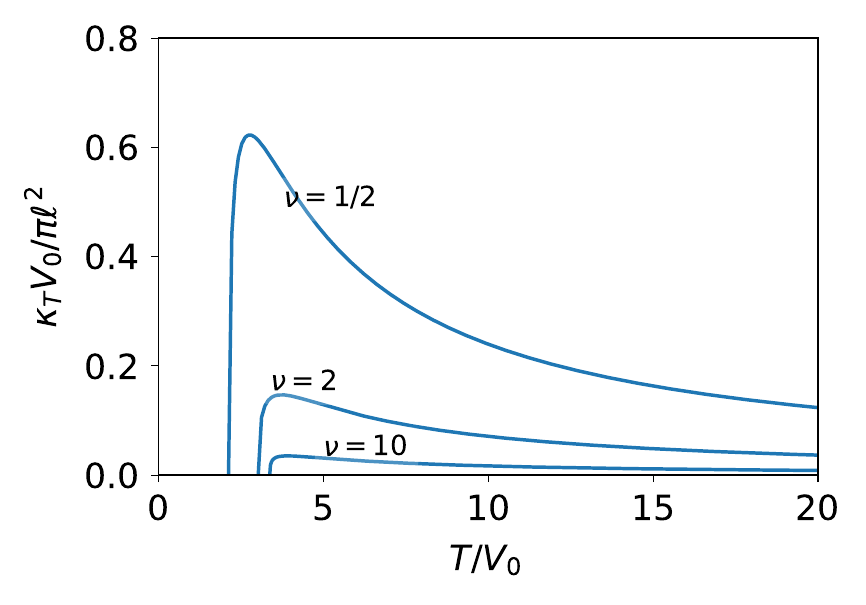}
\vspace*{-10pt}
\caption{Isothermal compressibility $\kappa_T$ as a function of dimensionless temperature $T/V_0$ for three different filling fractions $\nu = 1/2,2$, and $10$.}\label{fig:2}
\end{figure}

In the opposite limit of repulsive interactions at low temperatures, where $\beta V_m\gg 1$, the pair amplitudes \mbox{$A_m \to 0$} in Eq.~\eqref{eq:virialcontact} vanish, indicating the complete suppression of pairs with relative angular momentum $m$. Remarkably, the evolution from a non-interacting gas towards a strongly correlated state upon lowering the temperature can be studied even at this leading nontrivial order in the virial expansion by considering the isothermal compressibility, which reads
\begin{equation}\label{eq:compressbility}
\kappa_T=\frac{1}{n^2}\frac{d n}{d\mu}\Bigg|_T =\frac{1}{n^2}\frac{\beta}{\pi\ell^2} \left[b_1z+4b_2 z^2 + \clO\left(z^3\right)\right] \,.
\end{equation}
By eliminating the fugacity $z$ between Eqs.~\eqref{eq:virialfilling} and~\eqref{eq:compressbility}, it is possible to determine the dimensionless compressibility $\kappa_T V_0/(\pi\ell^2)$ as a function of temperature for different values of the filling~$\nu$. As shown in Fig.~\ref{fig:2}, this gives rise to a non-monotonic evolution towards states with small compressibility if the temperature is reduced at a fixed value of the short-range repulsion $V_0$. In particular, as expected on physical grounds, the crossover is more pronounced at small filling fractions, where mean-field theory no longer applies.

It is obvious that our extrapolation of the leading-order results for the pair amplitudes will not be quantitatively reliable at low temperatures, where the $\tilde{A}_m$ are no longer small and will also depend on the particle statistics. Yet, it is remarkable that the expected smooth crossover towards incompressible ground states, with a maximum of the compressibility $\kappa_T$ at some intermediate temperature of order $V_0$, is already captured by this approximation.

\section{Scale invariance and breathing mode in the LLL}\label{sec:scaleinvariance}

Important insights into incompressible fluid ground states of electrons in the LLL were obtained by considering a model where the repulsion at short distances completely dominates the interaction. For fermions, this is obtained by truncating the interaction Hamiltonian to $\hat{H}_{\intt}\to V_1\hat{P}_1$ with a single Haldane pseudopotential $V_1$~\cite{haldane83,trugman85}. In the context of ultracold gases, the corresponding reduced interaction $\hat{H}_{\intt}\to V_0\hat{P}_0$ for bosons is not just an idealized model but it provides a realistic description unless long-range forces become important like in dipolar gases~\cite{cooper05, cooper08}. For this type of interaction, which only affects pairs of particles with vanishing relative angular momentum $m=0$, the bosonic versions of the Laughlin wave functions for filling factors $\nu=1/2, 1/4 \dots$ are exact zero-energy eigenstates describing incompressible Bose fluids. Beyond numerical approaches, however, not much is known about the special features of this model, especially at filling fractions $\nu>1$ or at finite temperature. One question, in particular, that we will address in the following concerns the problem of what happens to the quantum anomaly, which arises due to the running of the coupling constant in 2D~\cite{olshanii10, hofmann12} in the absence of rotation. It turns out that within the LLL, the associated breaking of scale invariance appears in a rather different form, which is caused by the non-commutative nature of the guiding center coordinates.

In the absence of rotation, it has been shown by Pitaevskii and Rosch~\cite{pitaevskii97} that Bose gases in an isotropic 2D harmonic trap exhibit an infinite sequence of excited states with frequency $2\omega$. The existence of an infinite ladder of excited states may be derived in a purely algebraic manner by noting that the Hamiltonian in the presence of a trap can be written in the form
\begin{equation}\label{eq:H+C}
\hat{H}_{\omega} = \hat{H} + \omega^2\,\hat{C} \qquad \text{with} \qquad \hat{C}=\frac{1}{2}\int d^2 x \, \bfx^2 \, \hat{\rho}(\bfx)\,.
\end{equation}
Here, $\hat{\rho}(\bfx)$ is the mass density and $\hat{C}$ the generator of special conformal transformations~\cite{castin04,werner06,nishida07}. Now, for scale-invariant interactions, the commutator $i [\hat{H},\hat{D}]=2\hat{H}$ of the Hamiltonian with the generator $\hat{D}=- i \sum_i \bfr_i \cdot \nabla_{\rmr_i}$ of dilatations has the same form as in the non-interacting case. Using the commutators $i[\hat{H},\hat{C}]=\hbar^2\hat{D}$ and $i[\hat{D},\hat{C}]=2\hat{C}$, it is then straightforward to show that the operators defined by
\begin{equation}\label{eq:ladder}
\hat{L}_{\pm} \; = \frac{\hat{H}}{2\hbar\omega}-\frac{\omega}{2\hbar}\,\hat{C} \pm \frac{i}{2}\, \hat{D}
\end{equation}
act like raising and lowering operators for eigenstates with excitation energy $2n\,\hbar\omega$. These excited states correspond to breathing mode excitations, and the scale symmetry implies that their frequency is not affected by interactions. The full spectrum of the system in a harmonic trap thus separates into a set of conformal tower states separated by $2\hbar\omega$ that are connected by the raising and lowering operators $\hat{L}_\pm$. To ensure a bounded spectrum, there must be a special state (called a primary state) at the bottom of each tower that is annihilated by $\hat{L}_{-}\,\vert 0\rangle =0$. In particular, the exact ground state is such a primary state~\cite{nishida07,zwerger21}.

Due to the vanishing commutator $[\hat{L}_\pm, \hat{L}_z] = 0$, breathing mode excitations generated by the operator $\hat{L}_\pm$ are monopole excitations, i.e., they do not change the angular momentum quantum number of the state. Hence, a scale-invariant system placed in a rotating harmonic trap---which only affects the Hamiltonian as $\hat{H} \to \hat{H} - \Omega \hat{L}_z$---will still show breathing mode excitations at fixed frequency $2\omega$ irrespective of the rotation frequency $\Omega$, although the primary states themselves will depend on $\Omega$. However, this argument relies on the assumption that $i [\hat{H},\hat{D}]=2\hat{H}$ holds in the absence of rotation. This is not the case, however, for ultracold gases in 2D. Indeed, in order to obtain a finite value of the 2D scattering length $a_2$, the bare zero-range interaction $V(\bfx)=g_2(\Lambda)\delta^{(2)}(\bfx)$ needs to be regularized by a running coupling constant $g_2(\Lambda)=-2\pi\hbar^2/(m^*\ln{\Lambda a_2})$, cf. App.~\ref{app:twobody}. As was shown by Olshanii~et~al.~\cite{olshanii10} and by one of the present authors~\cite{hofmann12}, this leads to an anomalous contribution to the commutator in the form
\begin{equation}\label{eq:commutator-anomaly}
i \left[\hat{H},\hat{D}\right]=2\,\hat{H} +\frac{\partial \hat{H} }{\partial \ln{a_2}} = 2\,\hat{H} +\frac{\hbar^2}{4\pi m}\,\hat{C}_2,
\end{equation}
where $\hat{C}_2$ is the Tan contact operator introduced in Eq.~\eqref{eq:contact-definition}. The quantum anomaly gives rise to a shift of the breathing mode frequency away from the scale-invariant value $2\omega$, which has been observed in two-component Fermi gases near a Feshbach resonance~\cite{peppler18,holten18}.

Now, for motion within the LLL, the magnetic length provides an intrinsic scale that renders the zero-range interaction well defined without a cutoff. In explicit form, this may be derived by noting that, within the LLL, the interaction needs to be expressed in terms of the guiding center coordinates $\hat{\bfR}_j$ rather than those of the bare particle positions $\hat{\bfr}_j$ (here, $j$ labels the particle index). Indeed, it is the guiding center operators that commute with the gauge invariant velocity operators $\hat{\boldsymbol{\Pi}}_j$ which appear in the quenched kinetic energy of the LLL~\cite{haldane90}. Expressing the LLL projected Hamiltonian
\begin{equation}
\hat{H}_{\intt}^{\LLL} = \clP_{\LLL} \hat{H} \clP_{\LLL} =
\frac{1}{2} \int \frac{d^2q}{(2\pi)^2} V_\bfq f^2(\bfq) \sum_{j\,\neq\,l} e^{i \bfq \cdot \left(\hat\bfR_j - \hat\bfR_l\right)}
\end{equation}
in terms of the guiding center operators, the Fourier transform $V_\bfq f^2(\bfq)$ of the effective interaction contains a form factor $f^2(\bfq) =\exp{(-q^2\ell^2/4)}$~\cite{haldane90}. Formally, this arises from the LLL projection of the phase factor $e^{i \bfq \cdot \bfr_j}$ and it reflects the nontrivial algebra $[\hat{\bar{\rho}}_q,\hat{\bar{\rho}}_{q'}]=2 i \sin(\ell^2 (\bfq \times \bfq')) \hat{\bar{\rho}}_{q+q'}$ of the projected density operators $\hat{\bar{\rho}}_q = \sum_{i=1}^N e^{- i \bfq \cdot \hat\bfR_j}$ first noted by Girvin, MacDonald and Platzman~\cite{girvin86}. At large momenta $q\ell\gg 1$, the bare interaction potential $V_\bfq$ is thus suppressed by a Gaussian envelope $f^2(\bfq)$, which renders the interaction $V(\hat{\bfR})$ in terms of the guiding center separation $\hat{\bfR}$ finite without the need of a cutoff regulator. Specifically, the effective potential within the LLL associated with a bare contact interaction in 2D becomes
\begin{equation}
\tilde{V}(\hat\bfR) = g_2 \int \frac{d^2q}{(2\pi)^2} \, e^{i \bfq \cdot \hat\bfR} \, \exp{\left(- \frac{q^2\ell^2}{4}\right)} = \frac{g_2}{\pi \ell^2} \, \exp{\left(-\frac{\hat{R}^2}{\ell^2}\right)}.
\end{equation}
This is a finite and well defined potential for any fixed value of $g_2$ and it requires regularization only in the limit \mbox{$\ell\to 0$}, for which $\tilde{V}(\hat\bfR) \to g_2 \delta^{(2)}(\hat\bfR)$ reduces to a delta-function interaction. Superficially, therefore, it seems that the anomaly is gone in the LLL. This turns out to be incorrect, however, as we will show in the following.

To see this, we first note that the LLL projection of the generators $\hat{C}$ and $\hat{D}$ of special conformal transformations and of dilatations,
\begin{align}
\clP_{\LLL} \hat{C} \clP_{\LLL} &= \left(\hat{L}_z+\hat{N}\right)/(2\omega) \label{eq:LLLC}
\\
\clP_{\LLL} \hat{D} \clP_{\LLL} &= \hat{N}, \label{eq:LLLD}
\end{align}
reduce to the angular momentum operator and a pure phase factor, respectively. In particular, the dilatation in the LLL is trivial, reflecting the fact that a finite scale transformation of the particle coordinate always affects the cyclotron radius as well, which is fixed in a given Landau level. Now, as pointed out above, the proper dynamical degrees of freedom in the LLL are not the bare coordinates and the associated derivatives which enter the operators $\hat{C}$ and $\hat{D}$ but the guiding center operators $\hat{\bfR}_j$. Correspondingly, a dilatation in the LLL must be defined by $\hat{\bfR}_j\to\lambda \hat{\bfR}_j$ with some scale factor $\lambda$. Denoting the generator of a scale transformation of the guiding centers by $\hat{D}_R$, its commutator with the interaction Hamiltonian is then of the form
\begin{equation}\label{eq:commutatorDR}
i \left[\hat{H}_{\LLL}, \hat{D}_R\right] = \left(2 + \ell \frac{\partial}{\partial \ell}\right) \hat{H}_{\LLL},
\end{equation}
which is valid quite generally for any centrally symmetric interaction potential. As a result of the non-commutativity $[\hat{X}_j, \hat{Y}_l] = - i \ell^2 \delta_{jl}/2$ of the guiding center coordinates, therefore, scale invariance is violated in the LLL despite the fact that the coupling constant $g_2$ is no longer scale dependent. In particular, full scale invariance only arises in the classical limit $\ell \to 0$, where the guiding center coordinates and the particle coordinates coincide. A corresponding result holds for fermions in the LLL with a pure $V_1$-interaction of the form $V_1\nabla^2 \delta^{(2)}(\bfx)$. The anomalous commutator is then of the form $i [\hat{H}_{\LLL}, \hat{D}_R] = (4 + \ell \partial_\ell) \hat{H}_{\LLL}$.

A full discussion of the consequences of Eq.~\eqref{eq:commutatorDR} for the physics within the LLL is unfortunately beyond the scope of the present work. In the following, therefore, we focus on the question of what happens to the breathing mode frequency in a harmonic trap, which has served as an experimental signature for the breaking of scale invariance for 2D gases in the absence of rotation~\cite{peppler18,holten18}. It is important to note that this mode represents transitions between neighboring Landau levels. Indeed, the raising openrator $\hat{L}_{+}$ defined above acts in a trivial way on states restricted to the LLL, cf. Eqs.~\eqref{eq:LLLC} and~\eqref{eq:LLLD}. This point has been emphasized earlier by Watanabe~\cite{watanabe06}, who noted that the extent of the gas in the LLL is fixed by its angular momentum.
An excitation that changes the radius of the cloud thus always involves a change in angular momentum, such that monopole excitations like the breathing mode cannot exist solely within the LLL. In order to discuss the size $\Delta \omega$ of the breathing mode shift away from the scale-invariant value $2\omega$, it is necessary to determine the corrections beyond the scale-invariant mean-field equation of state. In a nonrotating BEC at low densities, this correction was found to give rise to a shift that is linear in $\tilde{g}_2$, $\Delta \omega/2\omega = \tilde{g}_2/16\pi$, with a numerically small prefactor~\cite{olshanii10,hofmann12}. On a microscopic level, these corrections to the low-density equation of state are determined from a self-consistent equation that links the chemical potential and the two-body T-matrix~\cite{fisher88}, and are thus tied to the logarithmic renormalization of the coupling constant, cf. App.~\ref{app:twobody}, which remains unchanged for rotating gases. For typical values of the interaction strength $\tilde{g}_2 \simeq 0.1$, the correction is numerically small and comparable to the second-order correction. Indeed, the second-order shift sets the leading order in Fermi gases and for few-body systems. An example of this is provided by the two-body calculation in App.~\ref{app:twobody}. Here, the relative pair energies for $m=0$ are $E_n = 2n \hbar \omega + V_0 + H_n V_0^2/2\hbar\omega + \clO(V_0^3)$, where $H_n$ is a harmonic number, which implies an anomalous breathing mode shift at second order in the interaction strength $\delta\omega/2\omega = V_0^2/4(n+1) \simeq \clO(\tilde{g}_2^2)$.

\section{Conclusion}\label{sec:conclusion}

In summary, we have discussed a parameterization of interacting bosons and fermions in the LLL in terms of Haldane's pair amplitudes $A_m$, which generalize a concept introduced by Tan for ultracold gases. It provides a description of translation- and rotation invariant states in the LLL beyond the usually considered case of incompressible phases. We illustrate this by determining the pair amplitudes in explicit form in terms of the Haldane pseudopotential parameters to leading order in the virial expansion. Remarkably, even this order provides a physically sensible description of bosons in the LLL at finite temperatures, which exhibit a maximum in the compressibility at temperatures of order $V_0$. As a concrete application of our results, we have shown that the data on the pair distribution function in the recent experiment~\cite{leonard22} with ultracold bosons in a periodically driven optical lattice allow to extract the two lowest Haldane pair amplitudes $A_0$ and $A_2$. They are consistent with a two-body state whose short-range correlations are close to that of a Laughlin state at $\nu=1/2$. Moreover, we have discussed what happens with the classical scale invariance for zero-range interactions in 2D in the presence of rotation. It has been shown that the logarithmic running of the coupling constant with momentum is absent in the LLL. Instead, a different quantum anomaly emerges which is associated with dilatations of the non-commutative guiding center operators. Finally, we have discussed the well known breathing mode in a harmonic trap, which now involves transitions between different Landau levels. We provide an estimate of the shift of its frequency away from the scale invariant result $2\omega$, which turns out to be of order $\tilde{g}_2^2\simeq 0.01$ for typical interactions. This provides a qualitative understanding of measurements of the radial breathing mode performed by Stock~et al.~\cite{stock04} in rotating Bose--Einstein condensates in a cigar-shaped trap, which do not detect a shift in the breathing mode frequency beyond effects of trap anharmonicity, consistent with theoretical predictions based on a scale-invariant equation of state~\cite{antezza07}.

There are a number of open problems which must be left for future study. First of all, it is clearly important to determine the shift of the breathing mode frequency in quantitative terms as a function of the dimensionless coupling constant $\tilde{g}_2$. On a more fundamental level, a deeper understanding of the special properties of truncated interactions like $\hat{H}_{\intt}=V_0\hat{P}_0$ or $\hat{H}_{\intt}=V_1\hat{P}_1$ for bosons or fermions, respectively, has been provided by Nguyen, Son and Wu~\cite{nguyen14}. They have shown that for zero-range interactions of this type, the time-reversal even and the odd response of incompressible ground states in the LLL at long wavelengths $q\ell\ll 1$ (but up to arbitrary order in the ratio $\omega/\Delta$ between frequency and the gap $\Delta$~\cite{golkar16}) are both determined by a single spectral function. As a specific consequence, the leading non-vanishing coefficient $s_4=(\clS-1)/8$ in the expansion $\bar{s}(\bfq)=s_4\,(q\ell)^4 +\ldots$ of the projected static structure factor is determined by the Wen-Zee shift $\clS$. It is an open problem to see whether further exact results can be derived from the approach in Ref.~\cite{nguyen14} about dynamical properties of either bosons or fermions in the LLL, whose interactions may be truncated to the leading nonvanishing contribution.

\section*{Conflicts of interest}
The authors have no conflict of interest to declare.

\section*{Acknowledgements}
We are grateful to Julian L\'eonard and Markus Greiner for their permission to use some of the data from their recent experiment~\cite{leonard22}. Moreover, W. Z. would like to take this opportunity to thank Jean Dalibard, to whom this issue of Comptes Rendus is dedicated, for the enjoyable collaboration on the review~\cite{bloch08} and for many insightful discussions over the last sixteen years, which have contributed substantially to his understanding of ultracold atoms.

\appendix

\section{Two-body problem in a rotating harmonic trap}\label{app:twobody}

In this appendix, we summarize results for the two-body problem in a rotating harmonic trap, which for contact interactions relate to well-known results in the absence of rotations~\cite{busch98,bloch08}.\linebreak In particular, as discussed in the main text, the solution shows explicitly that the cutoff-dependence in the strength of the delta-function interaction, which is necessary to regularize the problem in the absence of rotation, is absent in the LLL.

\begin{figure}[!htbp]
\centering
\includegraphics[scale=1.5]{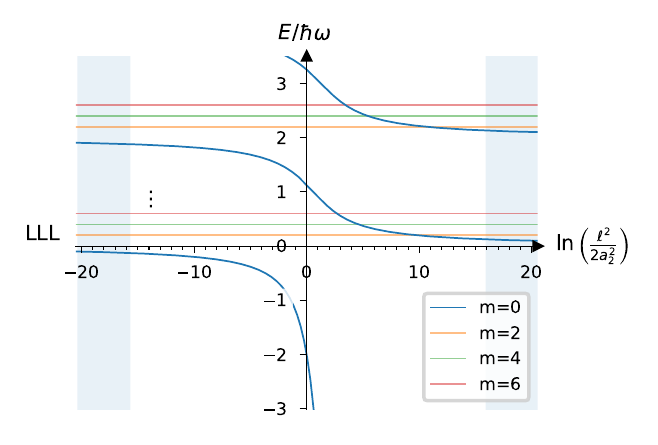}
\vspace*{-12pt}
\caption{Two-particle spectrum of 2D bosons with contact interactions in an isotropic harmonic trap as a function of interaction strength. Blue lines indicate the relative $m=0$ branches, which are affected by the interaction and independent of the rotation frequency. For illustration, we also include the first noninteracting branches with nonzero angular momentum for $\Omega/\omega = 0.9$. The blue shaded areas indicate the perturbative region.}\label{fig:3}
\end{figure}

The Hamiltonian for two particles in a rotating trap separates into a center-of-mass and a relative part:
\begin{align}
H_{2\rm{b}} &= H_{\com} + H_{\rel}
\end{align}
with
\begin{align}
H_{\rel} &= - \frac{\hbar^2 \nabla_r^2}{m^*} + \frac{m^*\omega^2}{4} r^2 - \Omega L_{z,r} + V_{\intt}(r) \label{eq:Hrel}
, \\
H_{\com} &= - \frac{\hbar^2 \nabla_R^2}{4m^*} + m^*\omega^2 R^2 - \Omega L_{z,R},
\end{align}
where $V_{\intt}(r)$ is the relative two-body interaction potential. It is already apparent from this form that for contact interactions, which select a relative zero-angular-momentum state, the trap rotation will not enter. Two-particle bound states are given by poles in the Green's function
\begin{align}\label{eq:poles}
\frac{1}{g_2} - G_E^{0}(\bfzo; \bfzo) &= 0,
\end{align}
where $G_E^{0}(\bfr; \bfr')$ is the noninteracting two-particle Green's function projected onto a center-of-mass eigenstate and evaluated at zero relative separation. Noninteracting eigenstates are
\begin{align}
\left\langle \bfr_1 \bfr_2\,\middle|\, n_Rj_R; nj \right\rangle &= \phi^R_{NJ}(\bfR) \phi^r_{nj}(\bfr)
\end{align}
with
\begin{align}
\phi^r_{nj}(z_r) &= (-1)^j \, \sqrt{\frac{1}{2 \pi \ell^2} \frac{n!}{j!}} \, \left(\frac{z_r}{\sqrt{2} \ell}\right)^{j-n} \, L_n^{j-n}\left(\frac{\bar{z}_r z_r}{2 \ell^2}\right) \, e^{-\bar{z}_r z_r/4\ell^2} \label{eq:phiCOM}
\\
\phi^R_{NJ}(Z) &= (-1)^j \sqrt{\frac{2}{\pi \ell^2} \frac{N!}{J!}} \, \left(\frac{\sqrt{2} Z}{\ell}\right)^{J-N} \, L_N^{J-N}\left(\frac{2 \bar{Z}Z}{\ell^2}\right) \, e^{-\bar{Z} Z/\ell^2} .\label{eq:phiREL}
\end{align}
For rapidly rotating traps with $\Omega \to \omega^-$, $N$ and $n$ are Landau level indices and $J$ and $j$ are the guiding center quantum numbers. The free Green's function is then given by
\begin{equation}\label{eq:greens}
\begin{split}
G_E^{0}(\bfzo; \bfzo) &= - \int_0^\infty ds \, \left\langle \bfzo\, \left| e^{\left(E - H_{\rel}\right) s} \right| \bfzo \right\rangle\\
&= - \frac{1}{2\pi \ell^2} \int_0^\infty ds \, e^{E s} \sum_{j=0}^\infty e^{- j 2 \hbar \omega s}\\
&= - \frac{1}{2\pi \ell^2} \int_0^\infty ds \, \frac{e^{E s}}{1 - e^{- 2 \hbar \omega s}}.
\end{split}
\end{equation}
As discussed, only the relative $m=0$ wave function contributes for the contact interaction. Performing the summation over all Landau levels in Eq.~\eqref{eq:greens} and introducing a short-time cutoff $m^*/\hbar \Lambda^2$ gives
\begin{equation}
G_E^{0}(\bfzo; \bfzo) = \frac{m^*}{4\pi \hbar^2} \left\{\gamma_E + \ln \left(\frac{2 m^* \hbar \omega}{\hbar^2 \Lambda^2}\right) + \psi_0\left(- \frac{E}{2 \hbar \omega}\right)\right\},
\end{equation}
where $\psi_0$ is the digamma function. We renormalize in the usual way by setting
\begin{equation}\label{eq:running}
g_2(\Lambda) = - \frac{2 \pi \hbar^2}{m^* \ln (a_2 \Lambda)},
\end{equation}
where $a_2$ is a 2D scattering length. The bound state condition now reads
\begin{equation}\label{eq:boundstate}
\ln \left(\frac{\hbar}{2 \omega m^* a_2^2}\right) = \ln \left(\frac{\ell^2}{2 a_2^2}\right) = \gamma_E + \psi_0\left(- \frac{E}{2 \hbar \omega}\right).
\end{equation}
This is of course identical to the bound state equation for contact-interacting particles in a 2D harmonic trap (with an energy shift by $-\hbar\omega$)~\cite{busch98}. A graphical sketch of the bound states is shown in Fig.~\ref{fig:3} as blue lines. We also include for illustration nonzero angular momentum states, which are not affected by the contact interaction and which form the Landau levels for $\Omega\to\omega^-$. There is one bound state branch that evolves from the attractive LLL, and which for $\omega \ll \hbar/m^*a_2^2$ has the standard bound state energy $E = - \hbar^2 e^{- \gamma_E}/m^* a_2^2$.

In order to restrict particle dynamics to the LLL as $\Omega \to \omega^-$, we require that the interaction shift is small compared to the LL spacing $2\hbar \omega$. This selects the perturbative regime, which is marked by the blue shaded area in Fig.~\ref{fig:3}. Here, the digamma function becomes $\gamma_E + \psi_0(- {E}/{2 \hbar \omega}) \to 2 \hbar\omega/E$, which gives the energy~\cite{busch98}
\begin{equation}
E = - \frac{2\hbar\omega}{\ln \left(2a_2^2/\ell^2\right)}.
\end{equation}
In this limit, the running of the coupling constant~\eqref{eq:running} is negligible and the 2D scattering length can be replaced by the unrenormalized perturbative coupling parameter $g_2$:
\begin{equation}\label{eq:EHartree}
E = \frac{g_2}{2\pi\ell^2} = \hbar \omega \frac{\tilde{g}_2}{2\pi} = V_0,
\end{equation}
which is precisely the LLL pseudopotential for a relative $m=0$ pair, as expected. The LLL limit $\hbar \omega/V_0 \to \infty$ then corresponds to the perturbative regime (with respect to the LL level spacing) that is manifestly scale invariant as established recently for a related case of 2D two-component Fermi gases~\cite{bekassy22}.

\section{Virial expansion in the lowest Landau level}\label{app:virial}

This appendix summarizes the virial expansion for states restricted to the LLL. We first discuss the virial expansion of the density operator
\begin{align}
\hat{n}(\bfr) &= \sum_{i=1}^N \delta^{(2)}\left(\bfr - \bfr_i\right),
\end{align}
which reads
\begin{align}
n(\bfr) &= \frac{\tr\left[e^{-\beta(H-\mu N)} \hat{n}(\bfr)\right]}{\clZ} = z n^{(1)} + z^2 \left(n^{(2)} - n^{(1)} Q_1\right) + {\clO}\left(z^3\right),
\end{align}
where ${\clZ} = e^{-\beta\Omega} = \tr [e^{- \beta (H-\mu N)}] = \sum_{N=0}^\infty Q_N z^N$ with $Q_N = \tr_N [e^{-\beta H}]$ is the partition function and $n^{(N)}(\bfr) = \tr_N[e^{-\beta H} \hat{n}(\bfr)]$. Here $\tr_N$ indicates the trace restricted to the $N$-particle sector. The virial coefficients then follow from the expansion for $\nu=(\pi \ell^2) n_2$ stated in Eq.~\eqref{eq:virialfilling} of the main text.

The single-particle trace for $n^{(1)}$ and $Q_1$ does not involve any interaction corrections and can be performed exactly using the single-particle states listed after Eq.~\eqref{eq:ham1} of the main text. The two-particle trace is evaluated using a basis of noninteracting two-particle states in a rotating trap listed in Eqs.~\eqref{eq:phiCOM} and~\eqref{eq:phiREL}, where we restrict the Hilbert space to the LLL level by fixing $N = n = 0$, in which case $m=j$ and $M=J$ with an energy of the two-particle state $\varepsilon_{M,m} = \hbar (\omega - \Omega) (M + m)$. 

Evaluating these few-body expectation values and taking the continuum limit $\beta \Delta \omega \to 0$ directly gives the virial expansion of the filling fraction~\eqref{eq:virialfilling} and the isothermal compressibility~\eqref{eq:compressbility}. Furthermore, the leading-order term in the expansion of the pair distribution function
\begin{align}
g^{(2)}(\bfr_1, \bfr_2) &= \frac{1}{n(\bfr_1) n(\bfr_2)} \left\langle \sum_{i \neq j} \delta^{(2)}(\bfr_1 - \bfr_i) \delta^{(2)}\left(\bfr_2 - \bfr_j\right) \right\rangle
\\
\intertext{is set by}
g^{(2)}(\bfr) &= \sum_m\,\!^{'} \frac{2}{m!} \left(\frac{r^2}{2 \ell^2}\right)^m e^{-\beta V_m} e^{-r^2/2 \ell^2} + {\clO}(z),
\end{align}
which immediately gives the contact amplitudes~\eqref{eq:virialcontact}. A similar calculation for the projected static structure factor gives
\begin{align}
\nu \bar{s}(q) &= \nu \bigl(1 \pm \nu \bigr) e^{- \frac{q^2\ell^2}{4}} + 4 \nu^2\sum_m\,\!^{'} e^{- \frac{q^2\ell^2}{2}} \left(e^{-\beta V_m}-1\right) \, L_m\left(\frac{q^2\ell^2}{2}\right) + {\clO}\left(z^3\right),
\end{align}
which is consistent with Eq.~\eqref{eq:barS} using the definition~\eqref{eq:virialAtilde}. It is straightforward to check that the leading order terms $\sim z$ satisfy the Ornstein-Zernike relation, which links the long-wavelength limit of the static structure factor to the isothermal compressibility,
\begin{align}
\bar{s}(\bfq=0) = n \kappa T,
\end{align}
where the density is given by~\eqref{eq:virialfilling} and the compressibility by~\eqref{eq:compressbility}.

\bibliographystyle{crunsrt}
\bibliography{crphys20221000}
\end{document}